\documentstyle[seceq,preprint,epsf]{jpsj}

\def\runtitle{Dynamical Properties of Chiral-Glass Order in Ceramic High-$T_c$  Superconductors}
\def\runauthor{Hikaru {\sc Kawamura}}

\title
{
Dynamical Properties of Chiral-Glass Order\\
in Ceramic High-$T_c$  Superconductors
}

\author
{ 
Hikaru {\sc Kawamura}
}

\inst
{
Department of Earth and Space Science, Faculty of Science,
Osaka University,Toyonaka 560-0043
}

\recdate
{
\today
}

\abst
{
There recently accumulated growing numerical and experimental evidence
that a novel glassy zero-field phase characterized by the spontaneously
broken time-reversal symmetry,
a chiral-glass phase, is
realized in certain ceramic high-$T_c$ superconductors.
Existence of frustration in zero external field, 
arising from the $d\/$-wave pairing symmetry of high-$T_c$
superconductors, is essential to realize this phase.
In this paper, we study the dynamical critical properties
of the chiral-glass transition by means of Monte Carlo simulations, 
based on a lattice {\it XY\/} model with finite screening.
We perform a dynamical scaling analysis, and the results are discussed
in conjunction with 
recent ac magnetic susceptibility and transport measurements
on high-$T_c$ ceramics.
}

\kword
{
high-$T_c$ superconductors, chiral glass, chirality,
vortex glass, frustration, critical dynamics, dynamical scaling 
}

\begin{document}
\sloppy
\maketitle

\section{Introduction}
Due to the enhanced effect of thermal fluctuations,
the problem of the phase diagram of cuprate 
high-$T_c$ superconductors becomes highly nontrivial
and has attracted much recent interest.
For example, in random type-II 
superconductors 
under applied magnetic fields,
the existence of novel glassy
thermodynamic phases distinct from the vortex-liquid phase, 
such as the Bragg-glass phase~\cite{rf:1} or the 
vortex-glass phase~\cite{rf:2}, have been discussed quite extensively.
In zero external field, by contrast, 
the only thermodynamic phase widely accepted 
to date is
the standard Meissner phase.
The $d$-wave nature of high-$T_c$ superconductors
appears to have little effect on  possible thermodynamic phases of 
bulk high-$T_c$ superconductors.

A few years ago, however, it was theoretically proposed that a new zero-field
phase, called a chiral-glass phase, might be realized in certain
{\em ceramic\/} or {\em granular\/} high-$T_c$ 
superconductors.~\cite{rf:3,rf:4,rf:5} 
This state is characterized by the spontaneously
broken $Z_2$ time-reversal symmetry with keeping the $U(1)$
gauge symmetry. The
order parameter is  a `chirality', 
representing the direction of
the local loop-supercurrent flowing over grains. 
In the chiral-glass state, circulation of 
local loop-supercurrents, spontaneously
generated even in zero field,  is frozen in time
in a spatially random manner. 
Frustration effect, which arises due to the random
distribution of the 
$\pi $ junctions with the negative Josephson coupling,   
is essential to
realize this phase. In this chiral-glass state, 
unlike in the Meissner or the
vortex-glass state, 
the phase of the condensate
is {\em not\/} ordered, even randomly, 
on sufficient long length and time scales: 
The  thermodynamic ordering occurs only in the loop-supercurrents, 
or in the chiralities. 

Numerical studies have given support to the existence of such a 
chiral-glass state both in the  presence and absence of screening
effect.~\cite{rf:3,rf:4,rf:5,rf:6,rf:7,rf:8,rf:9} In particular, 
static critical properties of the  chiral-glass transition
were studied  by means of Monte Carlo (MC) simulations by Kawamura and Li,
based on a lattice {\it XY\/} model in which finite screening effect
was taken into account.~\cite{rf:5} Here note that, since the length unit
associated with the intergranular ordering is the mean grain size
which is of order micron,  screening effect is generally 
non-negligible.
At the chiral-glass transition point, 
the nonlinear susceptibility was found to
diverge with  negative sign.
Indeed, this prediction was 
supported by subsequent ac susceptibility measurements on 
YBa$_2$Cu$_4$O$_8$ ceramics by Matsuura {\it et al\/}.~\cite{rf:10}

The purpose of the present paper is,
following the work of ref.5 on the static critical properties, 
to investigate the {\it dynamical\/}
critical
properties of the chiral-glass transition, and to discuss its implications
to  magnetic and transport measurements on high-$T_c$ ceramics.
A  quantity playing a central role in characterizing 
the dynamical critical properties is  the dynamical critical exponent $z$.
In order to numerically estimate $z$, 
we perform  dynamical (off-equilibrium) MC simulations on 
the model previously studied  in ref.5.
Equipped with both the dynamical and static properties of the model, 
we present a dynamical scaling analysis
for the magnetic response (ac susceptibility) 
and for the transport property
(resistivity). The results are  discussed in conjunction with some 
recent experimental results on high-$T_c$ ceramics.

\section{Model and Static Properties}

In this section, we explain the model and summarize its static critical
properties.~\cite{rf:5}

Regarding a ceramic sample consisting of many superconducting grains 
as an infinite network of Josephson-junction array, 
we consider
a three-dimensional lattice {\it XY\/} model where the phase variable is
coupled to fluctuating magnetic-field variable via a finite self-inductance
term.
The Hamiltonian is given in the dimensionless form by~\cite{rf:4,rf:5,rf:11}
\begin{eqnarray}
{\cal H}/J = - \sum _{<ij>} J_{ij}\cos (\theta _i-\theta _j-A_{ij})
+ \frac {1}{2{\cal L}}
\sum _p ({\mib \nabla} \times {\mib A})^2, 
\end{eqnarray}
where $\theta _i$ is the phase of the condensate of the grain
at the $i$-th site of a simple cubic lattice,
{\mib A}  the fluctuating gauge potential at each directed link
of the lattice, 
$J_{ij}$ the Josephson coupling
between the $i$-th and $j$-th grains, and $J$ is the typical coupling strength.
The lattice curl ${\mib \nabla} \times {\mib A}$
is the directed sum of $A_{ij}$'s around
a plaquette.
${\cal L}$ is the dimensionless self-inductance of 
a loop (an elementary plaquette),
while the mutual inductance between different loops
is neglected.
The first sum is taken over all nearest-neighbor pairs, while the
second sum is taken over all elementary plaquettes on the lattice.
Fluctuating  variables to be summed over are the phase variable,
$\theta _i$, at each site and the gauge variable, $A_{ij}$, at each
link.

The only source of quenched randomness
of the present model lies in  the Josephson
coupling $J_{ij}$. 
The $d$-wave nature of high-$T_c$ superconductors, which is vitally
important in realizing the chiral-glass phase at all, 
manifests itself in the form of the distribution of the Josephson
coupling $J_{ij}$.
In unconventional supercondutors with anisotropic
pairing symmetry such as $d$-wave superconductors, the
Josephson coupling between two superconducting grains could
be either positive ($0$-junction) or negative ($\pi $-junction),
depending on the relative orientation of the crystal.
Frustration arises even in zero external field from the random
distribution of both positive and negative  couplings, just 
as in case of spin glasses. 
The situation here should be contrasted to
the well-known vortex-glass (gauge-glass) problem. In the latter,
the Hamiotonian lacks the time-reversal symmetry 
due to external fields, while
frustration arises from  the external fields, 
not from  $J_{ij}$.
In the following, we assume for simplicity $J_{ij}$ to 
be an independent
random variable taking the values 
$1$ or $-1$ with equal probability ($\pm J$ or binary 
distribution), each
representing the $0$ and $\pi $  junctions.

The dimensionless self-inductance can be related to the 
bare Josephson penetration depth 
in units of lattice spacing, $\lambda_0$, by
\begin{eqnarray}
\lambda _0=1/\sqrt{{\cal L}},
\end{eqnarray}
Thus, larger inductance  corresponds to shorter
penetration depth with enhanced effects of screening.
In the limit ${\cal L}\rightarrow 0$ (or
$\lambda _0\rightarrow \infty $), the screening effect becomes negligible
and one recovers the standard $XY$ spin-glass Hamiltonian studied
in refs.6-9.

Note that the Hamiltonian (2.1)
possesses the $Z_2$ time-reversal symmetry, in addition to the $U(1)$
gauge symmetry. The chiral-glass state is characterized by the
spontaneous breaking of the discrete $Z_2$ time-reversal symmetry
with preserving the continuous $U(1)$ symmetry.
In the standard
vortex-glass problem,   the Hamiltonian has  the
$U(1)$ gauge symmetry only, and the ordered state (vortex-glass state)
is characterized by the
spontaneous breaking of this  continuous $U(1)$ gauge symmetry.

We define the local chirality 
at each plaquette by the gauge-invariant
quantity,
\begin{eqnarray}
\kappa_p=2^{-3/2}\sum_{<ij>}^p 
J_{ij}\sin (\theta_i-\theta_j-A_{ij}),
\end{eqnarray}
where the sum runs over a directed contour
along the sides of the plaquette $p$. 
Physically, the chirality
is a half ($\pi $) vortex, being proportional to the  
loop-supercurrent circulating around a  plaquette.
If the plaquette $p$ is frustrated, the local chirality
$\kappa _p$ tends to take a value around $\pm 1$, each sign
corresponding to the clockwise or counterclockwise circulating
supercurrent,
while if the plaquette is unfrustrated, 
it tends to take a value around zero. 
Note that the chirality is a pseudoscalar in the sense that it is
invariant under the global $U(1)$ gauge transformation, 
$\theta _i\rightarrow \theta _i+\Delta \theta,\ A_{ij}\rightarrow
A_{ij}$,
but changes its sign under the global $Z_2$ time-reversal transformation,
$\theta _i\rightarrow -\theta _i,\ A_{ij}\rightarrow
-A_{ij}$.

%\begin{figure}[bht]
%\begin{center}
%\noindent
%\figureheight{4cm}
%\epsfxsize=0.45\textwidth
%\epsfbox{dynamiccg_fig1.ps}
%\end{center}
%\caption{The temperature and size dependence of the Binder ratio of 
%the chirality, reproduced from ref.5b.}
%\label{fig:1}
%\end{figure}

Now, according to ref.5, we summarize the static critical properties
of the model.
%
%In Fig.1, we reproduce the temperature  dependence of the
%Binder ratio of the chirality $g_{{\rm CG}}$
%calculated for the $N=L\times L\times L$ lattices with $3\leq L\leq 10$
%via  equilibrium MC
%simulations. (Details  of the
%simulations  and the definition of  $g_{{\rm CG}}$
%have been given in ref.5.) 
%The bare penetration depth is set here equal to 
%$\lambda_0=1$.
%As can be seen from the figure, 
%the data of $g_{{\rm CG}}$ for 
%$L=3,4,6,8$ all cross at almost the same
%temperature $T\sim 0.28-0.29$, indicating
%the occurrence of a  finite-temperature
%chiral-glass transition at . In particular, 
%the data below $T_{{\rm CG}}$ show
%a rather clear fan out. 
%
The most extensive calculations were performed for the case
$\lambda_0=1$.
The Binder ratio of the chirality
was calculated 
for the $N=L\times L\times L$ lattices with $L=3,4,6,8$, which yielded a
clear crossing at a finite temperature. From this observation,
the chiral-glass transition temperature was estimated to be
$T_{{\rm CG}}=0.286\pm 0.01$ (temperature $T$ is
measured in units of $J$).
Via a finite-size scaling analysis of the chiral Binder ratio 
and of the chiral-glass
susceptibility,  static critical exponents of the chiral-glass
transition were estimated to be $\nu =1.3\pm 0.2$ and $\eta =
-0.2\pm 0.2$.  Other exponents were estimated via the standard scaling relation
as $\beta \simeq 0.5$ and $\gamma \simeq 2.9$, {\it etc\/}.
Monte Carlo data also clearly showed that the nonlinear susceptibility
exhibited a negative divergence at the chiral-glass transition.
As mentioned,
such a behavior was indeed observed
by ac susceptibility measurements.~\cite{rf:10}

\section{Dynamical Critical Properties}

In this section, we perform further MC
simulations
for the model (1.1), aimed at estimating the dynamical
critical exponent $z$ associated with the chiral-glass
transition. In contrast to the previous simulation of ref.5 
on the same model
where the system was completely thermalized,
we employ here an {\it off-equilibrium\/} Monte Carlo simulation where
the system is no longer equilibrated completely. Still, one can
extract the 
exponent associated with {\it equilibrium\/} critical dynamics
by controlling the waiting-time dependence of the results. More specifically,
we first quench the system from infinite temperature to the chiral-glass
transition temperature $T=T_{{\rm CG}}$, and  compute the subsequent 
temporal decay 
of the chirality autocorrelation 
function  defined by, 
\begin{eqnarray}
C_\kappa (t_w,t+t_w)=\frac{1}{3N} \sum_p [<\kappa _p(t_w)
\kappa _p(t+t_w)>],
\end{eqnarray}
where $t$ and $t_w$ are the observation and the
waiting times, respectively, $<\cdots >$ represents  the thermal average,
[$\cdots $] represents
the configurational average over the bond distribution, and
the sum runs over all $3N$ plaquettes on the lattice.
As long as one is in the 
so-called quasi-equilibrium regime, $1<<t<<t_w$, 
the decay
of  autocorrelations at the critical point
should obey the standard power-law,
\begin{eqnarray}
C_\kappa (t_w,t+t_w)\approx t^{-a},\ \ \ \ a =\frac{1+\eta }{2z},\ \ \ 
(1<<t<<t_w),
\end{eqnarray}
where the decay exponent $a$ is given by the dynamical
exponent $z$ and the static exponent $\eta $ as given above.
We have used here the standard
scaling relations and put $d=3$.
Thus, if one could well control the waiting-time dependence of
the data
in the quasi-equilibrium regime, 
one can estimate the equilibrium dynamical exponent
from off-equilibrium simulations.
Main advantage of this method is that one can deal 
with relatively large
lattices since one need not equilibrate the system completely.

We mainly simulate  $L=24$ lattices
with free boundary conditions, with varying the waiting time in the range
$5,000\leq t_w\leq 125,000$. Note that the lattice size studied here, 
$L=24$, is significantly
larger than the maximum lattice size, $L=8$, equilibrated
at $T=T_{{\rm CG}}$ in ref.5. 
Sample average is taken over 10
independent
bond realizations. For each realization, we perform four independent runs
with using different initial conditions and different sequences of
random numbers.
In order to check the possible size-dependence, we also take data
for $L=16$ lattices. 

The Hamiltonian (1.1) still has redundant degrees of freedom
associated with local gauge transformations,
which are usually fixed by the particular choice of the gauge. 
While static properties
do not depend on the  choice of the gauge, it is not entirely
clear whether  dynamical properties would not depend on it.
Thus, we employ two different gauges in our off-equilibrium simulations.
One is the
`temporal gauge'
in which the gauge-independent phase difference $\Psi_{ij}\equiv
\theta_i-\theta_j-A_{ij}$ is taken to be an independent variable,
and the other is
the Coulomb gauge  in which the divergence-free condition $\sum_\delta 
A_{i+\delta}=0$ is imposed at any site $i$. 
In the temporal gauge, we perform the standard Metropolis updating
successively on the $\Psi_{ij}$ variable at each link. In the Coulomb gauge,
we perform the  Metropolis updating
successively, first on the phase variable $\theta _i$ at each site, 
and then on the
gauge variables $A_{ij}$ {\it at each plaquette\/}: In the latter procedure,
in order to observe the local Coulomb-gauge condition,  we propose
a  type of MC trial which simultaneously shifts the four directed link
variables $A_{ij}$ around a plaquette by the same amount $A_{ij}\rightarrow 
A_{ij}+\delta A$.
We put $\lambda _0=1$ 
(or ${\cal L}=1$), 
and set the temperature to the chiral-glass transition temperature 
for this inductance, $T=T_{{\rm CG}}=0.286$,
which was determined by the previous equilibrium
simulation.~\cite{rf:5}

\begin{figure}[bht]
\begin{center}
%\figureheight{4cm}
\noindent
\epsfxsize=0.65\textwidth
\epsfbox{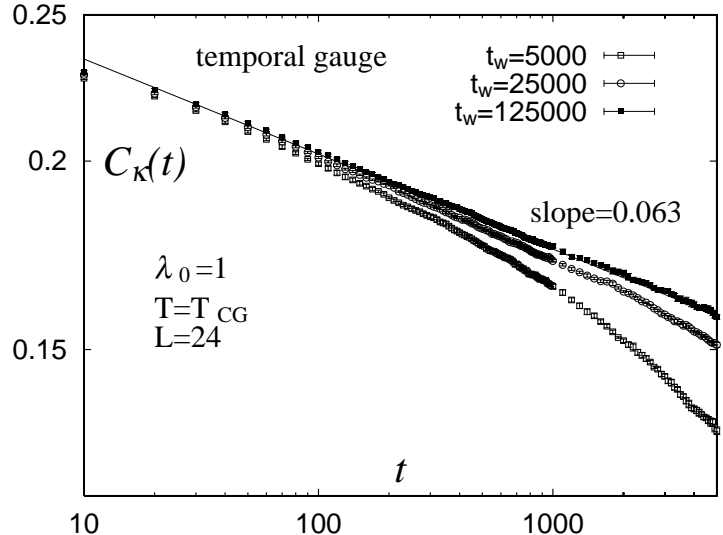}
\end{center}
\caption{A log-log plot of the observation 
time dependence of the chiral autocorrelation function
calculated in the temporal gauge
for several waiting times.
The bare penetration depth is  $\lambda _0=1$, and 
the temperature is set at the 
chiral-glass transition temperature,
$T=0.286$. The lattice size is $L=24$.} 
\label{fig:1}
\end{figure}

In Fig.1, we show on a log-log plot the MC time  
dependence of the chirality
autocorrelation function at $T=T_{{\rm CG}}$ calculated in the temporal
gauge.
As can be seen from the figure, the short-time data for 
longer waiting times
tend to lie on a common straight line with a slope equal to $a
=0.063\pm 0.006$. 

In order to check
consistency, we also calculate the spin-glass-type chirality 
autocorrelation
function defined by
\begin{eqnarray}
q^{(2)}_\kappa (t_w,t+t_w)=\left[\left<\left( \frac{1}{3N} 
\sum_p \kappa _p(t_w)\kappa _p(t+t_w)\right)^2 \right>\right].
\end{eqnarray}
At $T=T_{{\rm CG}}$, $q^{(2)}_\kappa$  in the quasi-equilibrium regime
is expected to 
decay asymptotically as
\begin{eqnarray}
q^{(2)}_\kappa (t_w,t+t_w)\approx t^{-2a},\ \ \ (1<<t<<t_w).
\end{eqnarray}

\begin{figure}[bht]
%\figureheight{4cm}
\begin{center}
\noindent
\epsfxsize=0.65\textwidth
\epsfbox{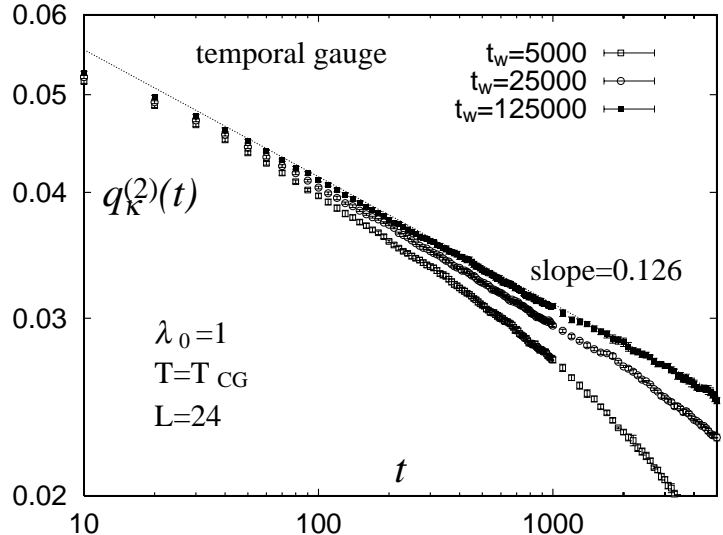}
\end{center}
\caption{A log-log plot of
the observation time dependence of the spin-glass-type
chiral autocorrelation function calculated in the temporal gauge
for several waiting times.
The bare penetration depth is  $\lambda _0=1$, and 
the temperature is set at the 
chiral-glass transition temperature,
$T=0.286$. The lattice size is $L=24$.}
\label{fig:2}
\end{figure}

In Fig.2, 
$q^{(2)}_\kappa $ calculated in the temporal gauge
is shown, which gives a slope
$2a=0.126\pm 0.012$,  consistently with the $a$ value obtained above from 
$C_\kappa$. We note that a 
similar analysis of $C_\kappa $ and $q^{(2)}_\kappa$ 
made for  smaller $L=16$ lattices has given the $a$ value close to this
indicating that the finite-size effect is negligible here.
Combining the estimate of $a$ with the previous estimate 
of $\eta =-0.2\pm 0.2$, the dynamical exponent  
$z$ is obtained as $z=6.3\pm 1.7$. 

To check the possible dependence on the choice of the gauge,
we repeat a similar calculation also
in the Coulomb gauge. As an example, 
the chiral autocorrelation function $C_\kappa $
calculated in the Coulomb
gauge is shown in Fig.3.
The obtained estimate 
$a=0.058\pm 0.006$ 
turns out to be slightly smaller than, but rather close  to  the 
value obtained in the temporal gauge. 

\begin{figure}[bht]
\begin{center}
\noindent
%\figureheight{4cm}
\epsfxsize=0.65\textwidth
\epsfbox{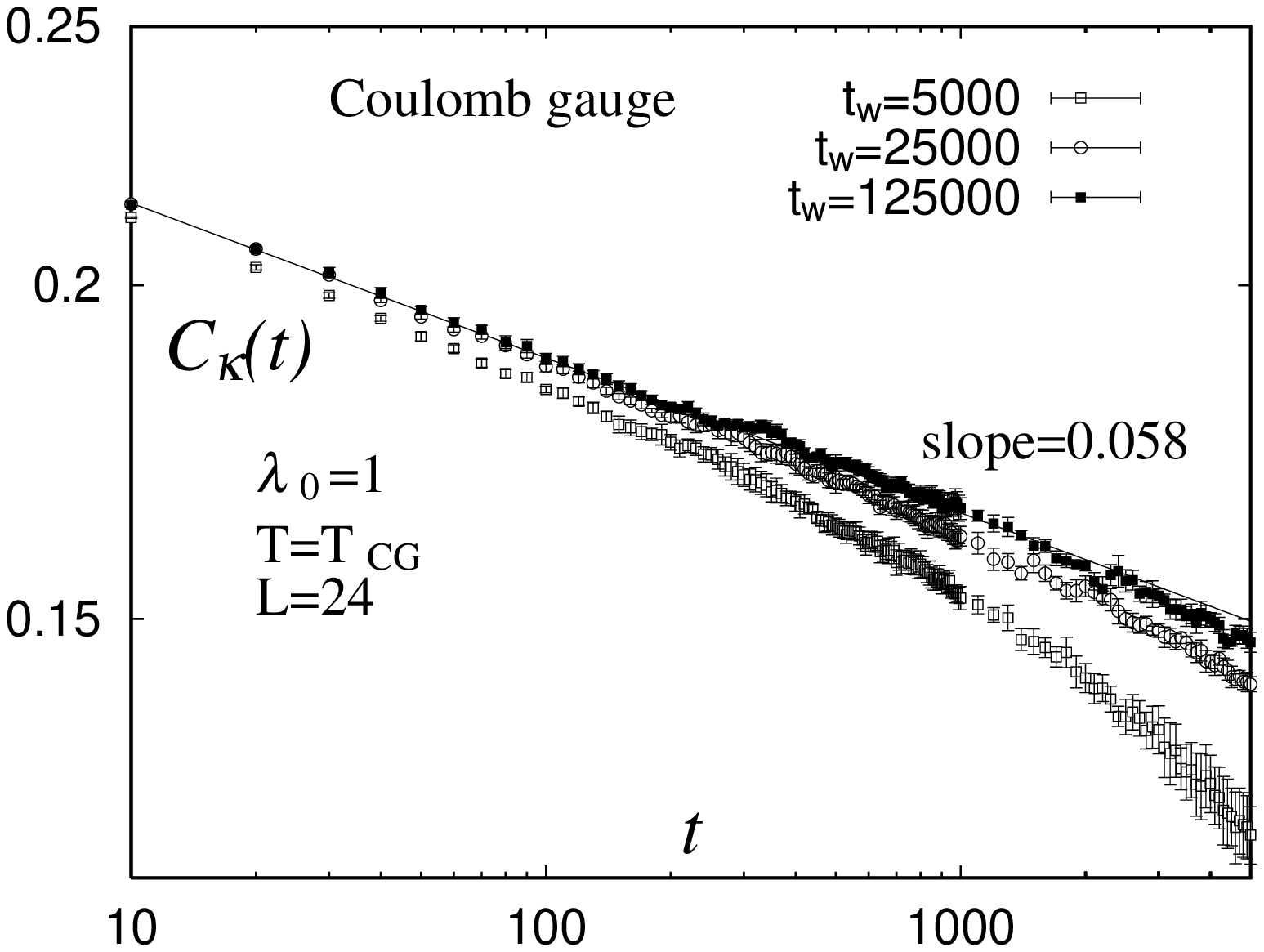}
\end{center}
\caption{A log-log plot of
the observation time dependence of the 
chiral autocorrelation function calculated in the Coulomb gauge
for several waiting times.
The bare penetration depth is  $\lambda _0=1$, and 
the temperature is set to the 
chiral-glass transition temperature,
$T=0.286$. The lattice size is $L=24$.}
\label{fig:3}
\end{figure}

One remark is to be added here. In the present simulation, we 
have implemented the standard Metropolis dynamics for
the Hamiltonian  in its original form,
{\it i.e.\/} in the  phase representation.
Instead, one can also 
employ the vortex representation of the
same Hamiltonian, {\it i.e.\/}, the dual version. In contrast to the
static critical properties, the dynamic critical properties
may well depend
on which representation is used. In fact,
in the numerical study of another model, {\it e.g.\/}, the
three-dimensional gauge-glass model,
%the gauge-glass model without screening, 
the phase and vortex representations
gave somewhat different $z$ values, namely, $z\simeq 3.1$ in the vortex
(dual) representation~\cite{rf:8}  which was considerably
smaller than $z=4.7\pm 0.7$ obtained in the 
phase representation.~\cite{rf:12} 
Hence, in the present model,
there still remains a possibility that the vortex representation yields
the $z$ value somewhat different from our present estimate 
obtained in the
phase representation. 
Unfortunately, we do not know at present which representation
is more appropriate in describing the real 
dynamics of experimental systems.

\section{Dynamical Scaling Analysis}

In this section, on the basis of the result obtained in the previous section,
we proceed to the dynamical scaling analysis both for the magnetic response
and for the transport.

\noindent
\subsection{Magnetic response}

First, we discuss the dynamical magnetic response
near the chiral-glass
transition, in particular the ac susceptibilities 
such as $\chi'' (\omega )$ and
$\chi' (\omega )$. 
Dynamical scaling analysis yields,
\begin{eqnarray}
\chi'' (\omega ,T)\approx \omega ^\frac{\beta
}{z\nu} \bar \chi'' (\frac { \omega }{t^{z\nu}})
\approx \omega ^{\frac{1+\eta}{2z}}\bar \chi'' (
\frac { \omega }{t^{z\nu}}),
\end{eqnarray}
and similarly for $\chi' (\omega ,T)$, 
where $z$  is the dynamical chiral-glass exponent determined
above as $z=6.3\pm 1.7$. Other exponents have been estimated numerically as
$z\nu \simeq 8.2$, $\beta \simeq 0.5$ and $\eta \simeq -0.2$ {\it etc\/}.

Experimentally, 
dynamic scaling analyses were made for the zero-field $\chi''$
data
of some high-T$_c$ ceramic superconductors such as LSCO~\cite{rf:13} and 
YBCO.~\cite{rf:14} In particular, 
Deguchi {\it et al\/} recently performed a dynamic
scaling analysis  of the 
zero-field $\chi''$ data of YBa$_2$Cu$_4$O$_8$ ceramics at its
intergranular transition point.~\cite{rf:14}
In this sample, the intergranular transition occurs at a temperature  much
below the intragranular transition temperature, the latter being the
bulk superconducting transition temperature of  single-crystal 
YBa$_2$Cu$_4$O$_8$. Since the effect of intragranular
transition is well separated from the intergranular one,
the sample is well suited to the study
of intergranular transition of interest here. 
Deguchi {\it et al\/} then found a reasonable scaling fit 
of the data, with 
the choice of critical
exponents, $z\nu\simeq 8$ and $\beta \simeq 0.57$, the values fairly close to
our present estimates $z\nu\simeq 8.2$ and $\beta \simeq 0.5$.

\noindent
\subsection{Transport property}

As mentioned, in the chiral-glass state,
the $U(1)$ gauge symmetry is {\em not\/} broken,
even randomly, in the
strict sense. This means that the  
phase of the condensate, $\theta $, remains
disordered   on sufficiently
long length and time scales.  Free motion of integer vortex 
lines is still possible
in the chiral-glass state where chiralities (half-vortices) 
sitting at
frustrated plaquettes are frozen.
A schematic picture showing such free motion of integer vortex-line
excitations in the background of frozen pattern of 
chiralities is given in Fig.4. One can see that 
free motion of integer-vortex lines of
either sign is possible without seriously destroying the 
freezing pattern of chiralities in the background.
In order to destroy the chiral-glass ordering in the background,
a chiral domain-wall-type excitation is necessary, which would
be responsible for the chiral-glass transition at $T=T_{{\rm CG}}$.
Thus, the chiral-glass state
should not be a true superconductor, with a small but nonvanishing
linear resistivity $\rho _L$ 
even at and
below $T_{{\rm CG}}$. Rough estimates of the residual $\rho _L$ was  
given in ref.5b.

\begin{figure}
%\figureheight{4cm}[bht]
\begin{center}
\noindent
\epsfxsize=0.5\textwidth
\epsfbox{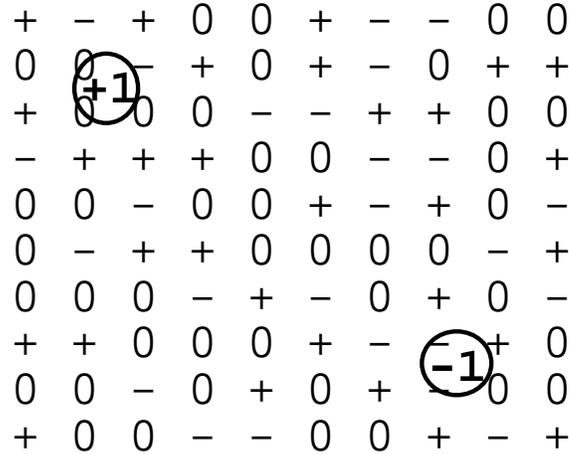}
\end{center}
\caption{Two-dimensional segment of the lattice showing 
thermally-activated integer vortex lines with 
vorticity $\pm 1$, wandering in the background
of a frozen pattern of chiralities in the chiral-glass state. 
Plus (+) and minus ($-$) chirality can be viewed as  half-vortices 
with vorticity $\pm 1/2$  at frustrated plaquettes,
while unfrustrated plaquettes are frozen into the 
zero-chirality ($0$) state. If one looks at 
a given frustrated plaquette frozen into  
the + chirality (or vorticity $+1/2$) state, for example, 
its vorticity 
occasionally becomes $+3/2$ or $-1/2$ when 
the thermally-activated integer vortex line of either sign, +1 
or -1, passes this plaquette.
Still, the long-time average of the vorticity at this plaquette is
equal to $+1/2$, showing that the free motion of integer vortex lines
is compatible with the long-range chiral-glass order.
}
\label{fig:4}
\end{figure}

Now, based on such a physical picture, we try 
to perform a dynamical scaling analysis of the transport
property of ceramic high-$T_c$ superconductors near the chiral-glass 
transition.
Suppose that, under the external current of density $j$, there 
occurs a voltage drop, or electric field of intensity $E$.
The above physical picture
suggests that the voltage drop comes from the two nearly independent
sources: One from the motion of integer vortex
lines, $E_v$, and the other from
the motion of chiral domain walls, $E_\kappa $. Namely,
one has
\begin{eqnarray}
E(j,T)=E_v(j,T)+E_\kappa (j,T).
\end{eqnarray}
The first part is expected to be essentially a regular part, while the
second part should obey the dynamic scaling law associated with the
chiral-glass transition, {\it i.e.\/},
\begin{eqnarray}
E_\kappa \approx \mid t\mid ^{(z+1)\nu }\bar E_\kappa (j/
\mid t\mid ^{2\nu }),
\end{eqnarray}
where $t$ is a reduced temperature $t\equiv (T-T_{{\rm CG}})
/T_{{\rm CG}}$, and
the spatial dimension has been set equal to $d=3$.
Standard analysis yields the following asymptotic behaviors
of the scaling function,~$^{2b}$
\begin{equation}
\bar E_\kappa (x)\approx 
\left\{ \begin{array}{ll}
ax, & (t>0),  \\ a'\exp[-bx^{-\mu }], & (t<0), \end{array}
\right. \ \ \ \ \ \ {\rm as}\ x\rightarrow 0,
\end{equation}
where $a$, $a'$ and $b$ are positive constants, and $\mu $ is an exponent
describing the chirality dynamics in the chiral-glass  state.

The linear 
resistivity $\rho _L$ can  also be 
written as a sum
of the two nearly independent contributions,
\begin{eqnarray}
\rho _L\equiv \left. \frac{{\rm d}E}{{\rm d}j}
\right | _{j=0}=
\rho _{L,v}+\rho _{L,\kappa}.
\end{eqnarray}
At and near the chiral-glass transition point $T=T_{{\rm CG}}$, the first term
$\rho _{L,v}(T)$ stays finite without prominent anomaly at  $T=T_{{\rm CG}}$, 
\begin{equation}
\rho _{L,v}(T) \approx \rho _0+ct+\cdots , 
\end{equation}
while the second term $\rho _{L,\kappa}(T)$ exhibits a singular behavior 
associated with the
chiral-glass transition, {\it i.e.\/}, one has from eq.(4.3)
\begin{equation}
\rho _{L,\kappa}(T) \approx 
\left\{ 
\begin{array}{ll} 0, & (t<0), \\ c't^{(z-1)\nu }, & (t>0). \end{array}
\right.
\end{equation}
Since $(z-1)\nu \simeq 6.9$ 
is a large positive number, the second term of eq.(4.5) vanishes
toward $T_{{\rm CG}}$ rather sharply, and the behavior of $\rho _L$
is dominated by the regular term $\rho _{L,v}(T)$. Thus,
the total linear resistivity remains finite at $T=T_{{\rm CG}}$.
The singular behavior borne by $\rho _{L,\kappa}(T)$ would be masked by the
regular term and  hardly detectable experimentally.

By contrast, a stronger anomaly could arise in the {\it nonlinear\/}
resistivity $\rho _{NL}$. If one considers the lowest-order nontrivial
one, it is again 
a sum of the two nearly independent contributions,
\begin{eqnarray}
\rho _{NL}\equiv \frac{1}{6} \left. \frac{{\rm d}^3E}{{\rm d}j^3}
\right | _{j=0}
=\rho _{NL,v}+\rho _{NL,\kappa}.
\end{eqnarray}
The first part is essentially a regular term, while
the second part can be written from eq.(4.3) in the scaling form,  
\begin{equation}
\rho _{NL,\kappa}(T)\approx 
\left\{
\begin{array}{ll} 0, & (t<0), \\ c''t^{(z-5)\nu }, & (t>0).
\end{array}
\right.
\end{equation}
Thus, the nonlinear resistivity
$\rho _{NL}(T)$ shows a stronger anomaly than the linear resistivity
$\rho _{L}(T)$. In particular,
if $z$ is smaller than five, the nonlinear resistivity
$\rho _{NL}(T)$ exhibits a positive {\it divergence\/} at $T=T_{{\rm CG}}$.
Then, the chiral-glass transition, even though it hardly manifests itself
in the linear resistivity, should clearly be detectable as a strong
anomaly in the nonlinear resistivity. Although
our present best estimate  obtained in the phase representation,
$z=6.3\pm 1.7$, 
comes slightly larger than five, it is not incompatible with a value
smaller than five within the errors. Furthermore, as discussed at the
end of \S 4, $z$ may possibly take a smaller
value in the 
vortex representation than in the phase representation.
Taking account of these uncertainties, there still exists a 
possibility that the value of $z$ relevant here is smaller than five, and
the nonlinear resistivity exhibits a real divergence at $T=T_{{\rm CG}}$.

Recently, Yamao {\it et al\/} measured via ac technique 
the linear and nonlinear resistivities of YBa$_2$Cu$_4$O$_8$ 
ceramics in zero external field.~\cite{rf:15} 
They observed a divergent behavior in the {\it nonlinear\/} resistivity 
just at the temperature where the nonlinear 
magnetic susceptibility showed a negative divergence and where the magnetic 
remanence set in. Meanwhile, the {\it linear\/} resistivity 
remained finite  without any appreciable anomaly there.
Such behaviors are hard to understand if one regards the observed transition
as the standard Meissner 
or the vortex-glass transition. Recall that
the linear resistivity should vanish in the Meissner or the vortex-glass
transition,
which is clearly at odds with the experimental finding of ref.15.
In contrast, the experimental result seems fully 
compatible with the chiral-glass picture discussed in this paper.

\section{Summary}

The dynamical critical properties of the chiral-glass
ordering were studied by means of Monte Carlo simulations and
the dynamical scaling analysis. 
We emphasize that the chiral-glass state is a new zero-field phase of
superconductors, being first made possible by the anisotropic nature of
the pairing symmetry of cuprate superconductors. 
In particular, we have shown that 
recent magnetic and transport measurements on YBCO
high-$T_c$ ceramics are consistent with the chiral-glass
picture. 

While in this paper  we have concentrated ourselves
on the equilibrium dynamical critical properties at and 
near the chiral-glass
transition, complete thermalization is practically impossible
deep into the chiral-glass state in real ceramics. There,  interesting 
off-equilibrium phenomena similar to those observed in spin glasses,
such as aging effects and  memory phenomena,
are certainly expected. For example, Papadopoulou {\it et al\/} recently
observed an aging effect in certain ceramic BSCCO sample at very weak fields,
~\cite{rf:16}
though the connection to the possible
chiral-glass order was not necessarily clear yet. For the future,
further theoretical and experimental
studies of  off-equilibrium dynamical properties of the chiral-glass
ordered state 
would be of much interest.

\section*{Acknowledgements}
The numerical calculation was performed on the FACOM VPP500
at the supercomputer center, Institute of Solid State Physics,
University of Tokyo. The author is thankful to
M.S. Li, M. Matsuura, T. Deguchi, M. Hagiwara,
M. Ocio, E. Vincent and P. Svedlindh
for valuable discussions.

\end{document}